# Near-field optical mode engineering-enabled freeform nonlocal metasurfaces

Zhongjun Jiang[1], Tianxiang Dai[2], Shuwei Guo[3], Soyaib Sohag[1], Yixuan Shao[2], Chenkai Mao[2], Andrea Alù[3,4*], Jonathan A. Fan[2*] and You Zhou[1*]

[1]Department of Physics and Optical Science, University of North Carolina, Charlotte, 28223
[2]Department of Electrical Engineering, Stanford University, Stanford, California, 94305, USA
[3]Photonics Initiative, Advanced Science Research Center, City University of New York, New York, NY, 10031, USA
[4]Physics Program, Graduate Center, City University of New York, New York, NY, 10016, USA

*Email: yzhou33@charlotte.edu, jonfan@stanford.edu, aalu@gc.cuny.edu

**Abstract**

Nanophotonic technologies inherently rely on tailoring light–matter interactions through the excitation and interference of deeply confined optical resonances. However, existing concepts in optical mode engineering remain heuristic and are challenging to extend towards complex and multi-functional resonant phenomena. Here, we introduce an inverse design framework that optimizes near-field distributions, ideally suited to tailor Mie-type modes within dielectric nanophotonic structures, and we demonstrate its powerful opportunities to facilitate the discovery of new classes of nonlocal metasurfaces. We show that freeform nonlocal metasurfaces supporting accidental bound states in the continuum can be readily optimized to tackle tailored illumination conditions, modal properties and quality factors. We further extend our approach to multifunctional and multipolar mode engineering, and experimentally demonstrate freeform planar nonlocal multi-wavelength and chiral metasurfaces. Our versatile and robust framework for freeform mode engineering has applications in a broad range of high quality-factor metasurface platforms relevant to sensing, nonlinear optics, optomechanics and quantum information processing.

Light-matter interactions bridging free-space waves and nanoscale resonant modes are crucial in the quest to engineer near- and far-field optical responses in nanophotonic technologies. In this quest, initial efforts have focused on the design of optical antennas with customized modes that support tailored scattering profiles and can collectively function as optical phased arrays[1,2]. These concepts have been later extended to Mie resonance engineering, involving the tailored excitation, coupling and interference of electric and magnetic multipolar resonances, at the basis of Kerker effect[3,4] and tailored bianisotropy,[5–8] and which are utilized in Huygen's metasurfaces[9–11], optical cloaks[12–14] and large angle metagratings[5,15–17]. Recent research efforts have applied these ideas to engineered nonlocalities, based on guided mode resonances and quasi-bound states in the continuum (BIC),[18,19] which can be tailored to enable far-field spectral filtering[20–22] and wavefront engineering with narrow band responses[23–26].

Despite the important role of resonances in this quest, a pathway to rationally tailor customized optical modes in structured media remains elusive due to the lack of precise analytical correlations between nanoscale geometry and near-field distributions. This observation is emblematic in the typical design process of nonlocal metasurfaces[27], which consists of a combination of physical intuition combined with numerical experiments. In a typical workflow, nanostructure geometries featuring non-radiating optical modes are proposed and identified using known physical relationships between nanostructures and modal symmetries. The layout symmetry is then carefully broken with spatially tailored perturbations, to enable weak coupling pathways between these highly confined modes and free-space radiation. Full-wave simulation sweeps are then used to empirically relate broken symmetries with nonlocal responses, building an alphabet of perturbations that allows the spatial structuring of nonlocal modes[26]. These approaches have been highly effective at developing the foundation of nonlocal metasurface research[28–32]. However, the complexity of local mode engineering and nonlocal responses that can be realized with these methods is limited, and it is challenging to extend these concepts to nonintuitive geometric shapes that support a full customization of multiple modes and functionalities within a single metasurface platform. It is not even clear whether there are fundamental limits of how many modal responses and functionalities can be packed within a single, ultrathin metasurface. In addition, the role of symmetries in the initial design makes it easier to tackle radiation towards high-symmetry points, and sophisticated dispersion engineering needs to be explored to rationally design metasurfaces with lower symmetry radiation[33].

In the following, we tackle this challenge by introducing a computational framework for freeform optimization of Mie-resonant metasurfaces that enables the explicit design of customized optical modes in the near-field. While inverse design techniques are typically used in metasurfaces to shape the far-field response, here we introduce a powerful adjoint optimization platform to shape the near-field modal distributions of nonlocal metasurfaces. Our approach bridges nanoscale mode engineering with freeform topology optimization, enabling the optimization of nonlocal metasurfaces within an exceptionally large design space and facilitating the discovery of new classes of metasurfaces with complex nanophotonic response utilizing nanoscale mode engineering. The workflow of our computational approach is presented in Figure 1a: first, given a desired near-field or far-field metasurface response, we build a framework of the optical mode physics that may be desirable (Figure 1a, left). We may tailor at will the mode profile, orientation, quality factor ($Q$-factor), wavelength and complex amplitude. The interactions between the modes and far-field radiation can be described using coupled mode theory (Figure 1a, center), which provides quantitative guidelines for specifying the objectives for metasurface optimization. For this study, we consider a simplified coupled mode theory picture that neglects intermodal coupling, though our concepts can be extended quite straightforwardly to include such mode interactions. To tailor the coupling between free-space waves and the desired nanoscale modes, we utilize an adjoint variables method (AVM)[34,35] adapted to the near-field, in which forward and adjoint simulations utilize a combination of near-field and far-field excitation sources (Figure 1a, right). Our study uses of a basic local gradient descent optimizer, though the local gradients calculated using AVM can ultimately be used in conjunction with a wide range of local and global optimization algorithms[17,36,37].

Our freeform optimization strategy, in which low performing geometric layouts evolve towards high $Q$-factor structures with desired modal profiles, supports distinctive features compared to conventional metasurface optimization approaches. Our platform is ideally suited for full-wave solvers and therefore fully accounts for and exploits the complex relationship between nanoscale freeform shape and optical mode properties without approximations. It does not need to assume or enforce high symmetries pertaining to the photonic nanostructures and incident waves, and it is therefore particularly useful at discovering new classes of accidental BIC structures, which represent a broad family of nonlocal metasurfaces challenging to identify through heuristic designs. Our platform also readily extends to devices hosting multiple multipolar resonances by the use of multi-objective optimization, and it can fully tailor wavelengths and $Q$-factors of each one of these modes, their complex amplitudes and spatial position in the device.

To illustrate the basic concept with a simple model system, we consider the design of an ultrathin, nonlocal, periodic dielectric metasurface that produces a single narrowband Fano-type far-field lineshape. We select an operating wavelength of 1500 nm with a period of 750 nm along the $x$- and $y$-directions, and we choose silicon as the dielectric material due to its low loss, high refractive index and compatibility with conventional Mie-type nanophotonic structures[38,39]. The device is excited with a normally incident plane wave, and we specify that this wave couples to an out-of-plane magnetic dipole Mie mode ($m_z$) to produce a Fano-type lineshape (Figure 2a). Under these conditions, the $m_z$ Mie mode is supported in highly subwavelength-scale film thicknesses, and has been identified in prior quasi-BIC-based metasurface demonstrations as a powerful tool to design efficient dielectric metasurfaces[21,24,27,29,40]. To ensure that only this Mie-type mode is supported in the metasurface, the film thickness is limited to 150 nm, suppressing higher-order multipolar resonances. Furthermore, the metasurface unit cell boundaries are specified to be air, which suppresses the formation of delocalized mode profiles spanning multiple unit cells.

A Figure of Merit (FoM)[31,32] is defined to maximize the complex amplitude of the desired $m_z$ Mie fields in the metasurface, given the desired incident far-field excitation source. A precise AVM setup for our problem therefore involves the forward simulation source to be a normally incident plane wave and the adjoint simulation source to be a current distribution profile capturing the $m_z$ Mie mode profile. Interestingly, we have found that it is possible to simplify our FoM to the maximization of the real part of complex amplitude $\text{Re}(H_z)$ at a point centered within the silicon metasurface unit cell, and to use the excitation of a point dipole $m_z$ source with a profile mimicking the $m_z$ Mie mode when performing adjoint simulations. Our use of a point dipole source in the adjoint simulations is effective because our nonlocal metasurface system exhibits an enhanced Mie mode optical density of states, and as the photonic Mie mode dielectric structures form during optimization, coupling between the point source and emergent Mie structure leads to predominantly Mie mode-based near-field electromagnetic field profiles. Our simple target function requires a single dipolar point adjoint source, but more complex adjoint sources can be generally used to capture higher-order or more complex Mie modal profiles.

Our optimization algorithm utilizes a two-part AVM-based optimization pipeline. First, pixel-based topology optimization is performed to identify metasurface topologies that roughly capture the desired coupling between the far-field source and near-field modes. Second, we fine tune the modal properties and Q-factors using AVM-based boundary optimization. A challenge posed by the optimization of high Q-factor photonic devices is the extreme sensitivity of the device properties to geometric perturbations. To address this challenge, we introduce a neuro-parameterization scheme to describe the metasurface layout, in which a neural network encodes analytic relationships between position and device layout. Such a scheme circumvents spatial resolution limits posed by pixel-based AVM design concepts by specifying layout features with unlimited spatial resolution. It also introduces new ways to introduce constraints important to experimental fabrication, such as feature size and curvature constraints, by framing constraints in the form of loss function engineering during network training and geometry updating. A more detailed discussion of this scheme can be found in our recent work[41].

The optimization trajectory tracking the FoM is presented in Figure 2b and shows three parts. First, the design is initialized as a uniform grayscale permittivity profile and a pixel-based AVM optimization at the operating wavelength is performed with continuous grayscale dielectric values to identify a promising device topology. Over the course of iterations, the FoM consistently increases and is ultimately enhanced by five-orders-of-magnitude compared to the starting FoM (Figure 2b, blue line). Second, the device evolves into a binarized freeform structure consisting of Si and air, and the magnetic field distribution within the structured media gradually transforms into a localized hotspot indicating the emergence of a high-Q magnetic resonance (Figure 2b, inset). Third, the FoM is fine-tuned using AVM-based boundary optimization with our neuro-parameterization scheme (Figure 2b, orange line). The detailed tuning of Q-factor as a function of the out-of-plane magnetic field intensity $|H_z|^2$ is shown in Figure 2c and shows a linear trend consistent with the relation $FE^2 \propto Q$, where FE denotes the local electromagnetic field. This trend is consistent with those known for critically coupled, lossless single mode systems[42]. The insets show the gradual geometric modifications of the high-resolution features leading to varying Q-factors, including enhanced Q-factors with values reaching $10^4$.

The transmission spectrum of our metasurface with a Q-factor of 448 is shown in Figure 2d and features a narrowband Fano resonance dip within a broadband transmission window, which is typical of the interference between radiative and nonradiative modes. To confirm the excitation of the $m_z$ Mie mode in the metasurface, we performed a multipolar decomposition of the metasurface near-fields from the current density distributions induced in the nanoscale resonators using the open-source software MENP[43]. The multipolar decomposition as a function of wavelength is shown in Figure. 2e and it reveals the excitation of a dominating magnetic dipole resonance and no additional modes. The magnetic field distribution inside the meta-atom at the resonant peak (Figure 2e, inset) further confirms the presence of a strong magnetic Mie mode resonance within the metasurface.

The nonlocal metasurfaces produced by our optimization scheme support quasi-BIC modes that differ from the widely studied symmetry-protected BICs[27]. The latter require symmetry constraints pertaining to the photonic nanostructures and the incident wave, typically emerge at the center of Brillouin zone, and are supported in highly symmetric arrays operating under normal incidence. In contrast, our approach facilitates accidental BIC formation purely from structural engineering[44,45], enabling new classes of nonlocal metasurfaces featuring asymmetric geometries and illumination conditions. As a demonstration, we use structures from Figure 2 as starting points and design a series of nonlocal metasurfaces that support strong accidental BIC responses for different oblique incidence angles (Figure 3a). As shown in Figure 3b, the Q-factors of the optimized metasurfaces may be consistently pushed above high values with minimal shift in the resonance frequency, which contrasts with the strong angular dispersion of quasi-BIC mode designed at a single angle (Supplementary Section 1). To confirm that the freeform-designed BICs are accidental and not symmetry protected, for the device optimized for normal incidence, we perform a shape interpolation between the freeform structure and the one of a symmetric donut structure, which is known to produce a symmetry-protected BIC[27]. As the interpolation ratio increases, the corresponding Q-factors decrease during the intermediate stages of shape interpolation, reaching a minimum at an intermediate interpolation ratio value (Figure 3c). This nonmonotonic trend highlights a transition between the accidental and symmetry-protected BIC schemes.

Our computational optimization framework can be readily extended to multifunctional metasurfaces, including the multiplexed excitation of various Mie modes with distinct properties within a single unit cell. As a first demonstration, we design a multi-wavelength non-local freeform metasurface that supports a pair of dipole resonances, each operating at distinct wavelengths and featuring distinct mode symmetries and resonant properties. The metasurface is designed on a 150 nm thick silicon layer, arranged in a periodic array with a lattice constant of 870 nm. The optimization setup is shown in the schematic in Figure 4a and shows the specification of two Mie-type modes in a periodic meta-array, an in-plane electric dipole ($p_x$) and an out-of-plane magnetic dipole ($m_z$) at the wavelengths of 1.3 μm and 1.365 μm, respectively. The two modes exhibit distinct field orientations relative to metasurface plane.

To co-optimize the two modes, we define a composite FoM within the metasurface as the sum of the ED and MD mode intensities, $|E_x(\lambda_1)|^2 + |H_z(\lambda_2)|^2$. The optimization trajectory of the mode intensities (log scale) at the two target wavelengths is shown in Figure 4b, indicating a consistent increase in FoM over the course of optimization. The transmission spectrum and the top-view permittivity profile of the optimized metasurface is presented in Figure 4c and shows two narrow transmission dips featuring different linewidths. To identify the mode characteristics of the excited photonic Mie modes, we perform multipolar decomposition of the metasurface scattering response and identify strong transverse electric ($p_x$) and vertical magnetic ($m_z$) modes at the transmission dip wavelengths (Figure 4d). The corresponding field distributions at the two peak wavelengths (Figure 4d, insets) further confirm strong electric and magnetic resonances hosted within the metasurface. More detailed results on the scattering cross sections of all other directional components can be found in Supplementary Section 2.

We experimentally validate the design by fabricating the optimized metasurface within a 150 nm-thick polycrystalline silicon film on a fused silica substrate. The metasurface patterns are defined using electron beam lithography and reactive ion etching. Figure 4e presents the top-view scanning electron microscope (SEM) images of the meta-atoms, showing well-defined geometric features consistent with the design. To measure the transmission spectrum, the sample is illuminated with a supercontinuum laser and imaged through a magnification system consisting of a near-infrared objective and a tube lens, and the transmitted radiation is collected by a spectrometer. The measured transmission spectrum (Figure 4f) confirms the presence of the two target dipole modes.

As a third demonstration, we designed a freeform metasurface that utilizes spectrally overlapped multi-mode resonances to exhibit planar chiral non-local responses[40,46], specifically spin-selective responses that are exclusively induced by circularly polarized light of a specific handedness (Figure 5a). Single layer metamaterial systems supporting strong chiroptical responses generally require the electric and magnetic fields to be co-linear and spatially overlapped within the dielectric medium[47]. To achieve these criteria, we use multi-objective optimization to specify spectrally and spatially overlapping $E_z$ and $H_z$ dipole modes that support co-linear electric and magnetic field components (Figure 5b). We specifically define the multivariate FoM to be $|E_z||H_z|$ at a point within in the metasurface, and the FoM is specified to be maximized only when the metasurface is illuminated by left-hand polarized light (LCP) incidence.

The metasurface is designed for a wavelength of 1460 nm and a period of 750 nm. The silicon layer thickness is specified to be 450 nm, and it is relatively thicker compared to prior demonstrations to break the radiation symmetry in the forward and backward directions. A top view of the optimized chiral metasurface is shown in Figure 5c (left), and the corresponding near-field $|E_z|, |H_z|$ distributions under LCP (Figure 5c, middle) and RCP (Figure 5c, right) illumination show strongly selective Mie mode excitations under LCP illumination. The simulated transmission spectra under LCP and RCP illumination are shown Figure 5d and show a 94% circular dichroism (CD) response at the design wavelength. A multipolar decomposition of the chiral metasurface reveals the excitation of co-linear ED and MD modes that are spectrally overlapped at the designed wavelength (see details in Supplementary Section 3). The designed chiral metasurface exhibits spin-selective conversion of circularly polarized input, where near-unity conversion occurs exclusively under LCP incidence. Detailed co- and cross-polarized transmission spectra are provided in Supplementary Section 4.

We experimentally validate the design by patterning and etching a 450 nm-thick silicon film on a fused silica substrate, and the top-view SEM image of the fabricated device is shown in Figure 5e. The left inset provides a zoomed-in top view of the nanoscale features, revealing well-defined curvilinear geometries that match the design. The right inset shows a tilted SEM image of the silicon pattern, showing smooth, vertical sidewalls indicative of high-quality silicon etching. The measured transmission spectra for the two CP illuminations (Figure 5f) show measured spectral line shapes that match well with the simulation. We attribute the reduced CD in the experimental device to fabrication imperfections and the slight off-normal incidence of the laser beam, as the nonlocal modes are sensitive to the phase matching conditions (see details of spectra analysis under oblique incident angles in Supplementary Section 5). Further enhancements can be achieved by imposing more stringent feature size constraints to mitigate sensitivity to fabrication imperfections[17,48–50].

In summary, in this work we have introduced and demonstrated a near-field inverse design framework to realize freeform nonlocal metasurfaces through the explicit engineering of optical modes. Our approach accounts for and exploits the complex interplay between nanoscale freeform shapes and optical near-fields, which enables full customization of optical modes in both spatial and spectral domains. The large design space offered by topology optimization can be readily extended to devices hosting multipolar and multifunctional resonances. We anticipate many future extensions of this work. One is the extension of our concepts beyond single-layer media to multilayer nonlocal metasurfaces[51–55], which can lead to qualitatively new regimes of mode multiplexing capabilities. Another involves incorporating spatial perturbations into engineered nonlocality to achieve spatial and momentum light control[25,56]. Another opportunity explores the utilization of faster electromagnetic solvers and optimizers[57,58], which may address current computational bottlenecks in throughput and speed. On the application front, we anticipate that our ability to customize optical near-fields has the potential to impact many application domains including molecular sensing, where conventional nonlocal metasurfaces are limited due to the location of hotspots within the metasurface nanostructures and where our platform can be used to define customized hotspots in near-field regions outside of the metasurface nanostructures. Multifunctional hotspot engineering also has applications in nonlinear optics[28,59], optomechanics[60,61], and quantum emission enhancement[62] and photochemistry[63,64], where strong and tailored light-matter interactions are required.

## Data Availability

The data that support the plots within this paper and other findings of this study are available from the corresponding authors upon reasonable request.

## Acknowledgments

Y.Z. acknowledges support from the UNC Charlotte Faculty Research Grant, the Center for Metamaterials, and UNC Charlotte start-up funds. J.F. acknowledges support from the Samsung and the National Science Foundation. A.A. acknowledges support from Simons Foundation and Air Force Office of Scientific Research.

## Contributions

Y.Z., A.A., and J.F. developed the idea. Z.J., T.D., Y.S., and Y.Z. conducted the modeling and theoretical analysis. S.G. and S.S. fabricated the samples. C.M. provided the silicon thin film. Z.J. and S.G. performed the experimental measurements and data analysis. Y.Z., A.A., and J.F. wrote the manuscript with input from all authors. The project was supervised by Y.Z., A.A., and J.F.

## Competing interest

The authors declare no competing interests.

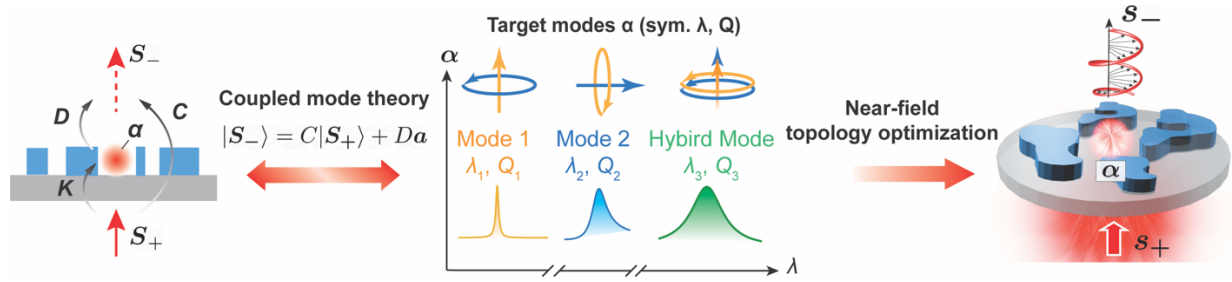

**Fig. 1| Mode engineering framework.** The process starts with identification of optical modes $\boldsymbol{a}$ that transform the incoming wave $\boldsymbol{S_+}$ into a desired outgoing wave $\boldsymbol{S_-}$ based on coupled mode theory (left). The modes are defined as design objectives (center) and are optimized through near-field topology optimization (right).

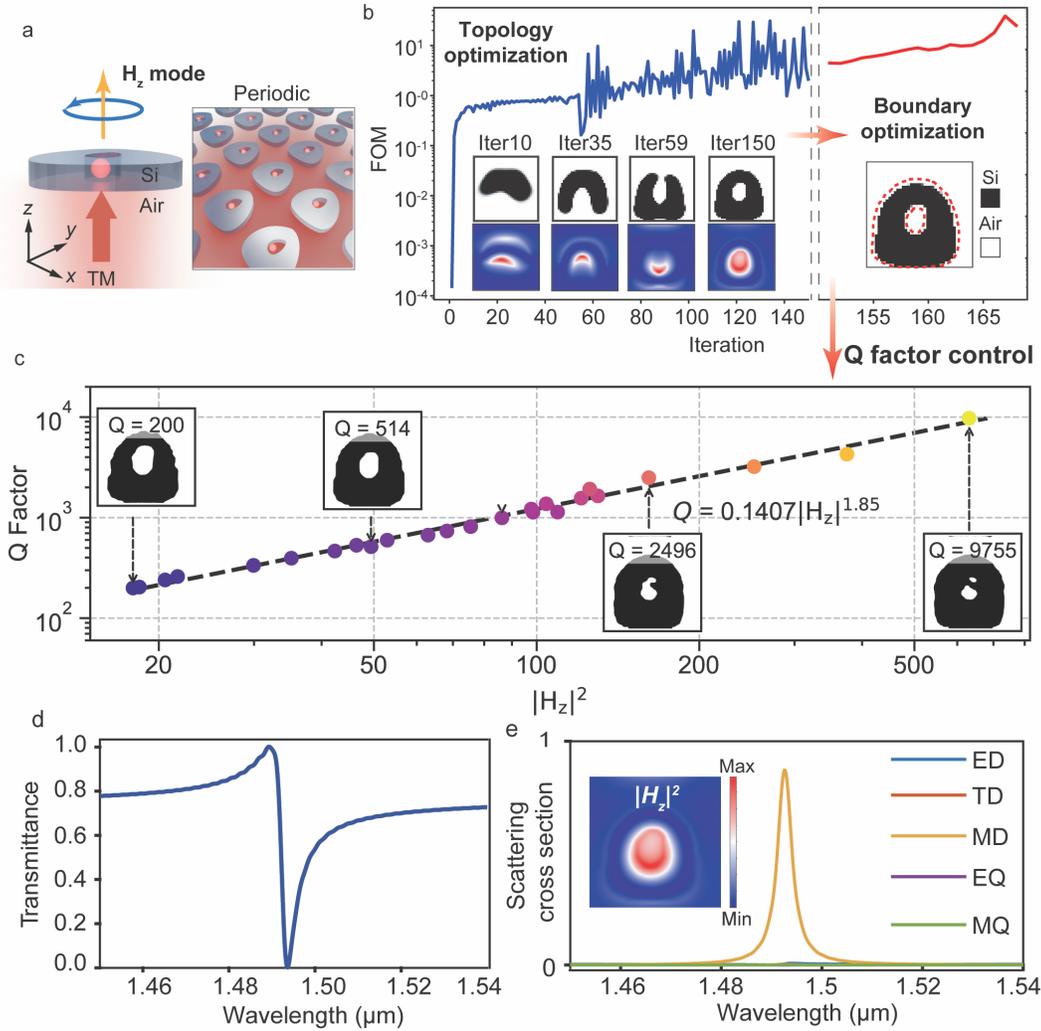

**Fig. 2| Freeform nonlocal quasi-BIC metasurfaces.** (a) Schematic of a nonlocal metasurface supporting a quasi-BIC mode featuring a vertical magnetic dipole ($m_z$) Mie mode. (b) Optimization trajectory showing the enhancement of the FoM over the course of topology and boundary optimization. Insets: structural and magnetic near-field evolution within a single unit cell. (c) Q-factor engineering using neuro-parameterized boundary optimization. Insets: unit cell structures for differing target Q-factor values. (d) Transmission spectrum of the optimized Metasurface with a Q-factor of 448. (e) Multipolar decomposition of the near-field at resonance. Inset: top-view of the magnetic near-field distribution, showing a clear magnetic dipole field profile.

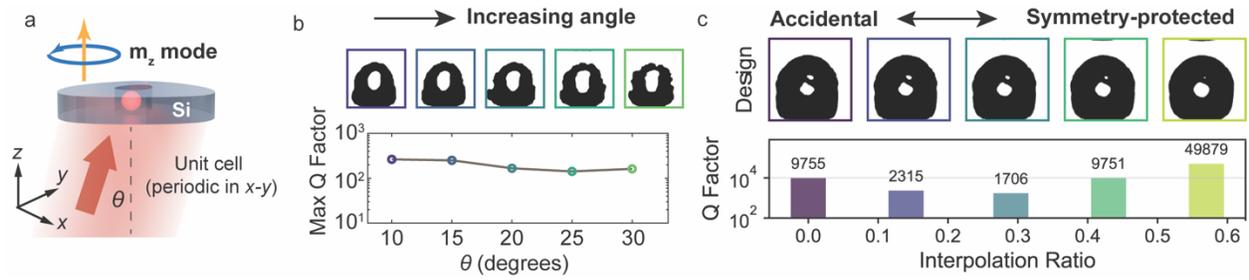

**Fig. 3| Accidental quasi-bound state in the continuum (quasi-BIC) design under oblique incidence.** (a) Schematic of a nonlocal metasurface supporting accidental BICs under oblique incidence. (b) Top: Designed unit cells for different off-axis incident angles ranging from 5 to 30 degrees. Bottom: Q-factors of optimized metasurfaces designed for different off-axis incident angles. (c) Shape interpolation of the highest Q-factor device from Figure 2c and a symmetric donut supporting a symmetry-protected BIC mode. Top: geometric evolution of the unit cell with increasing interpolation ratio. Bottom: Q-factor as a function of interpolation ratio, showing a nonmonotonic transition between the two BIC regimes.

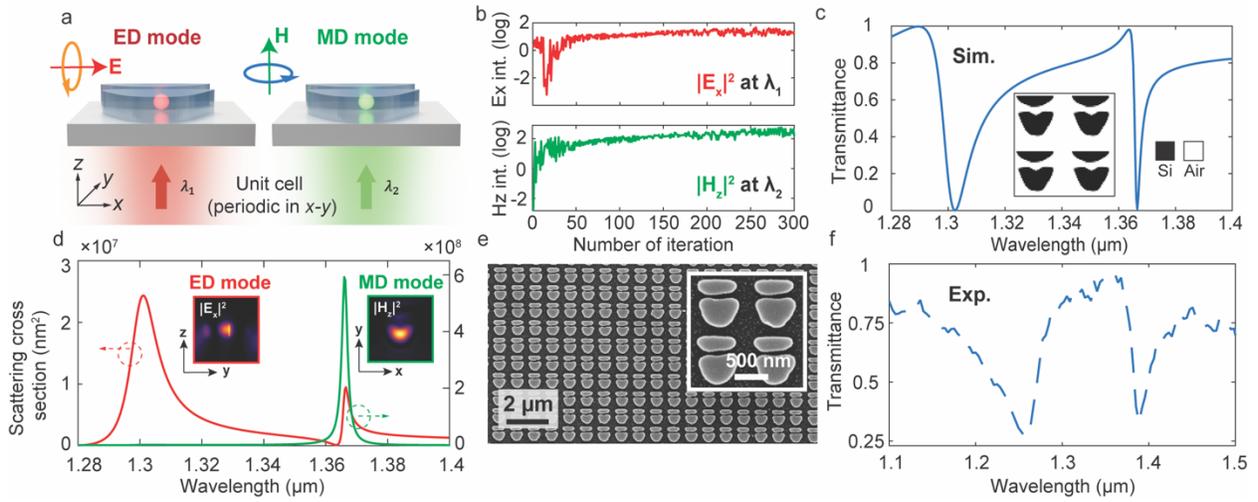

**Fig. 4 | Multispectral freeform mode engineering.** (a) Design implementation for a multiwavelength mode-engineered nonlocal metasurface. The device is designed to support an in-plane electric dipole (ED) resonance and an out-of-plane magnetic dipole (MD) mode at two different wavelengths. (b) Optimization trajectory showing the log-scale increase of electric and magnetic field intensities at the two target wavelengths. (c) Simulated transmission spectrum of the optimized structure. Inset: top view of the freeform device layout. (d) Multipolar decomposition of the near-fields. Insets: field profiles of the two target modes. (e) Scanning electron microscope image of the fabricated device. (f) Measured experimental transmission spectrum.

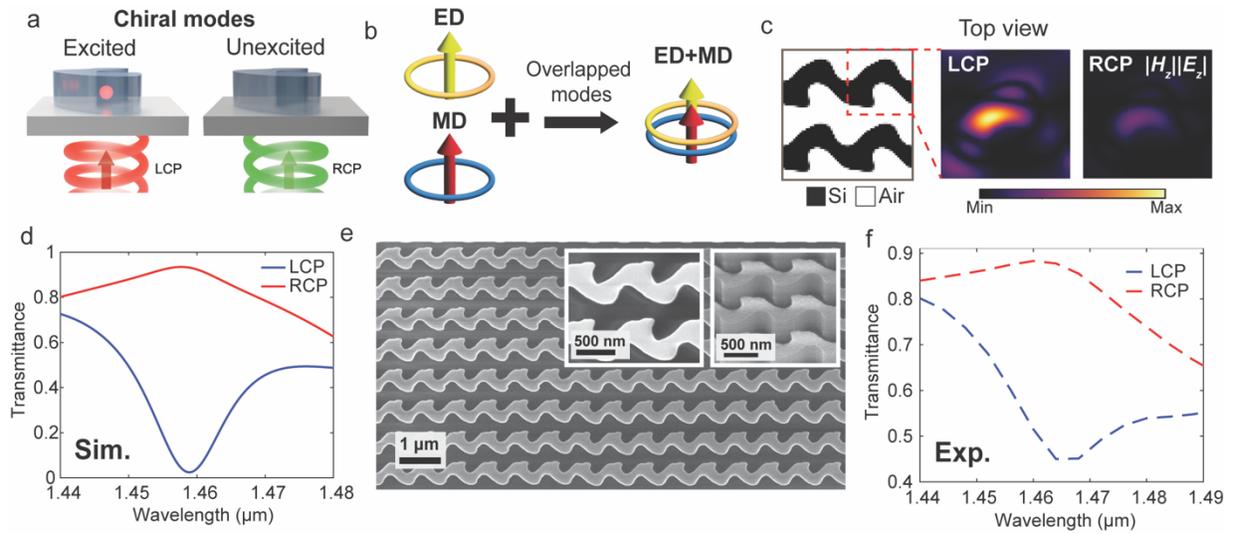

**Fig. 5 | Multi-resonant chiral mode engineering.** (a) Design implementation showing that the chiral modes are excited exclusively under left-hand polarized light (LCP) incidence and remain unexcited for right-hand polarized light (RCP) incidence. (b) Multi-resonant mode design concept involving a pair of spectrally overlapped electric dipole and magnetic dipole modes. (c) Top view of the optimized metasurface (left), with the corresponding $|E_z||H_z|$ near-field distributions for LCP (center) and RCP (right) illumination. (d) Simulated transmission spectra for LCP and RCP incidence. (e) Scanning electron microscope images of the fabricated device. (f) Measured experimental transmission spectra for LCP and RCP incidence.